# Monetary Policy and Stock Market

Kian Tehranian

*Department of Economics and Finance, Queen Mary University of London, UK*

*August 2020*
## Abstract

This paper assesses the link between central bank's policy rate, inflation rate and output gap through Taylor rule equation in both United States and United Kingdom from 1990 to 2020. Also, it analyses the relationship between monetary policy and asset price volatility using an augmented Taylor rule. According to the literature, there has been a discussion about the utility of using asset prices to evaluate central bank monetary policy decisions.

First, I derive the equation coefficients and examine the stability of the relationship over the shocking period. Test the model with actual data to see its robustness. I add asset price to the equation in the next step, and then test the relationship by Normality, Newey-West, and GMM estimator tests.

Lastly, I conduct comparison between USA and UK results to find out which country's policy decisions can be explained better through Taylor rule.

Corresponding author. Tel.: +1 (424)-407-6624
*E-mail address:* tehraniank@g.ucla.edu

I would like to thank Professor Konstantinos Theodoridis for his most valuable comments and suggestions on this paper.
1

# Introduction

There are two most commonly used tools for controlling economic activity, monetary and fiscal policy. Fiscal policy is determined by government and the aim is to control economic growth and inflation by adjusting the government tax rate and the government spending level. It does, however, affect the government budget and debt level. On the other hand, central bank is known as an independent national authority regulates banking, monetary policy, and financial services such as economic research. The central bank is owned by the government but independent from government finance department or ministry, and functions as a banker, advisor, and agent to the government. Throughout the economy, it plays a crucial role which directly and indirectly affects everyone in financial transactions. Its main objective is to stabilize the currency of the country, low unemployment rate and control inflation. The road between monetary policy and the stabilization operates in a number of ways. The six primary sources are Interest rate, exchange rate, equity, bank lending, wealth and balance sheet.

Some channels aim to have a greater economic influence. The channels will eventually work together, though. Monetary policy triggers changes in the interest rate, the exchange rate and the value of the financial assets, which then influence consumption expenditure, investment expenditure and net export, all resulting in changes in the gross domestic product (GDP) and other macroeconomic factors. Changes in GDP lead in changes in unemployment rates, price levels and inflation. A rise in GDP, for instance, lowers unemployment but raises inflation. A reduction in GDP is causing unemployment to increase and inflation to decrease. The most commonly used tool is the interest rate, which operates by increasing both consumption and investment expenditure. Lower interest rates usually contribute to higher investment and consumption expenditure.

In this paper I analyze the effectiveness of the Taylor rule, which is known as a tool to link the Central Bank's instrument (short-term nominal interest rate) to the inflation rate, output gap and asset price volatility in both the United States and the United Kingdom. From 1993 Taylor introduced the theory as a framework for evaluating monetary policy and behavior of central bank system. Ever since economists used the rule to analyze the decisions of policy makers. The popularity of the rule resets on some features.

First, it is straightforward as it ties the interest rate directly to economic conditions, as captured by inflation and output gap. There is, however, an implicit link in targeted rules where central banks seek to reduce deviations from a target. Plus, in comparison with inflation forecast targeting, it does not need a prediction model. Setting current inflation and output gap is sufficient.

In addition, the Taylor rule has defined the monetary policy actions with precision. For example, Gelach and Schnabel (1998) found that the Taylor rule in the 1990s defines average nominal short-term interest rates in the euro area with a coefficient of 0.5 on output gap and 1.5 on inflation. Ultimately, simplicity and performance in tracking nominal short-term interest rates and evaluating monetary policy decisions illustrate the major usage of Taylor rule by economists and private-sector analysts.

Taylor rule, on the other hand, has two key disadvantages from the prospective monetary authorities. There is small number of variables in the feedback list, so it is too limiting. Generally speaking, it is irrational for central banks not to use other information such as the exchange rate, other asset prices, credit aggregates, etc. in support of price stability. In addition, the methods may not be resilient to systemic shifts in the economy. The efficient coefficients would be complicated functions of the economic model's structural parameters and the desires of the policy maker. For example, changes in the structure of the economy generally lead to changes in the efficient coefficient. Central banks aren't willing to implement such a simple rule, according to the above reasons. Central banks need to be able to flexibly adjust policy in response to new data and economic structural changes.

But for three purposes, Simple Taylor rule, government guidance, may be helpful. Firstly, it can be used internally as a benchmark to test the decisions taken by policy makers based on different information presented. The existence of Taylor rule as a benchmark gives consistency to the workers at the central bank to justify their research deviates from the one indicated by benchmark. And policymakers use the Taylor rule to determine a potential interest rate range, but they are pragmatic in exercising their discretion and allowing the interest rate to deviate from the amount indicated by the rule. Secondly, the rule should be used as a simple mechanism to justify policy decisions for the general population, and the rule understood to the population should help to minimize confusion regarding future monetary policy and deter macroeconomic volatility.

As described above, the advantages of using the Taylor rule will depend on how effective the rule's ability to control inflation and output gap is in modifying the economy structure. Of example, if the policy deviates regularly and continuously from the benchmark and so the rule has to be updated frequently, thus the benefit of using benchmark would vanish.

Historical policy assessment has been seen as an attempt to clarify the relationship between policy decisions and economic outcomes, and thus to assess whether policy action was appropriate in terms of timing, magnitude and direction. Although the Taylor rule serves as a useful tool for interpreting past policy decisions and errors, adoption of the Taylor rule is not enough to avoid further policy errors.

The monetary policy aim of the central bank is both market stability and sustainable economic development. Recently, policymakers and academic economists have been attempting to resolve the issue of whether, in addition to price stabilization, central bank should still recognize asset price stability (Federal Reserve Chairman Alan Greenspan 1996). So, some are of the opinion that in their decisions central banks should consider volatility of asset prices, especially the stock market. One explanation for this is that solvency and liquidity issues may be triggered by significant and market-wide shifts in asset prices. For an example, the distribution of resources can be disordered if financial markets are reversely influenced by unpredictable price fluctuations. Another explanation may be that the high volatility in stock prices leaves investors unsure about potential profits, and therefore may induce an economic slump or even initiate one by itself. Because of the reverse trend in acquisitions and transactions (Hu (1195); Choudhry (2003)), companies and businesses may delay their ongoing consumer buying until the future seems to be more certain. Therefore, instability in the stock market will contribute to greater macroeconomic uncertainty in terms of output and volatility in inflation.

Another purpose of this paper is to determine the impact of financial market uncertainty on central bank policy decisions by changing interest rates. Therefore, it is important to find acceptable participants of the stock market in the United States and the United Kingdom. Since the S&P500 and FTSE100 comprise the 500 and 100 biggest market capitalization companies in the US and UK respectively, they can be a fair indicator of asset price volatility.

**Literature Review**

Capistran (2008) reports that policymakers are more interested in high than low inflation. Branch (2014) indicates that higher inflation and output forecasts make policymakers more cautious, as well as demonstrating that uncertainty affects policy decisions, higher uncertainty leads to more passive monetary policy.

There is no consensus among economists on whether monetary policy and asset price are related or not. Economists give two replies. One party argues, Bernanke and Gertler (1999, 2001), the right monetary strategy is that it sets inflation rates regardless of whether or not asset prices are perceived. In other words, asset price volatility does not affect target inflation for the central bank. Yet another group, Cecchetti (1998), believes that information about asset prices can improve economic performance.

Bernanke and Gertler (1999) suggest that central bank would pay no heed to asset price inflation, because reasonable interest rates will stabilize asset prices. They also note (1999, 18) that stabilizing asset prices is troublesome because it is unclear whether a shift in asset values is triggered by fundamental factors, non-fundamental factors or both.

Bullard and Schaling (2002), use the macroeconomic model to evaluate the inflation, output and asset price targeting. They believe that a strategy that responds to asset prices can be detrimental because it can conflict with inflation minimization and fluctuations in output. It also claims that a strategy of managing asset prices will lead to more unpredictable fluctuations than they happen, if asset prices were not taken into account. Targeting asset prices is weakening monetary policy efficiency and undoubtedly doing significant harm. Goodfriend (2003) had the same idea that the direction and size of the correlation between the volatility of asset prices and the short-term interest rate is not stable. Consequently, the correct course and scale of the interest rate relative to asset values is in fact challenging. Finally, Filardo (2000, 2001) concluded that if asset prices are high, monetary policy would respond to asset prices, otherwise it would be better to stay neutral. Filardo and Fair (2000) states that there is no evidence that by paying attention to asset prices through the macroeconomic model, Fed can boost economic stability. They demonstrate the adverse impacts of capital loss following a stock market collapse overshadow the Fed's positive consequences by reducing the interest rate after the crisis.

By comparison, Cecchetti (1998) says monetary policy should consider asset prices, as stated earlier. The explanation for this is that policy makers tend to trade off output variability for price volatility because it is difficult to keep them steady. Also, Cecchetti and Krause (2000) asses the relationship between considerable movement in financial structure of many regions and conclude the movements have caused the stability of economic development and low inflation. In 2003, he argues the policy reactions to stock price changes should be different according to its reason, for example growth in earnings and profits. Bernance and Gertler (2001 p. 257) believe that the shock cycle for non-fundamental asset markets is entirely different, although the equations are almost the same. As a result, policymakers know that the observable fluctuations are not due to underlying factors and can thus boost economic efficiency by reacting to shifts in market markets. Besides, Rigobon and Sack (2003) use the methodology of heteroscedasticity to analyze the response of monetary policy to stock markets. They claim that there is a clear response to changes in the stock market, with swings of 5 per cent in the S&P500 Index raising the possibility of 25 basis point tightening or easing by around half. They break down all daily and weekly fluctuations in interest rates and stock market values, thereby concluding that short-term stock market moves push interest rates in the same direction as stock price changes.

## Data Set

The variables that need to be assessed in Taylor rule are real Gross Domestic Product (GDP), Consumer Price Index (CPI), Nominal Short-term Interest Rate and Stock Market Index.

Short term nominal interest rate is the expense of borrowing funds, usually calculated as the loan's annual amount. It's the premium banks spend for their short-term deposits.  It is the principal instrument of the central bank to control the economy. Therefore, policy makers use various approaches to determine optimal interest levels according to the target level of the economy indicators. Basically, as the interest rate increases, so does the inflation and stock market.

Law leaders and analysts use Gross Domestic Product (GDP) to calculate the value of all domestically produced goods and services. GDP is essential, since it shows an economy's size and well-being. The rise in GDP should be seen as a good indicator of overall economic health. There are three main methods of calculating a country's GDP. They are both winding up with the same number of results.

This is the form of spending, which measures all the various forms of spending such as: consumption, investment, government expenditure and net exports.

Second method is Earnings. In this method all factors of payment should be added together. The factor contributions are made up of profits, labor returns and capital returns.

Lastly, the system of production, which measures the overall value of all manufactured products minus the value of intermediate goods.

The authorities mostly use CPI as an indicator of inflation rates. It essentially tests shift in the average market basket price index for households. Therefore, all the goods and services that people purchase will be in the market basket including food, accommodation, transportation, medical treatment, education, etc. The CPI Index is calculated by dividing market basket costs in a given year by market basket costs in the base year, and multiplying by 100. Therefore, inflation rate can be driven by the percentage change in CPI Index from base year to the given year.

The stock index is a stock market tracking index which is developed by individual stocks and allows investors to compare current prices with past values and quantify performance. There are plenty methods to calculate stock index.

Full form of market capitalization is the line up of the market capitalization of the company. In the United States the S&P500 index uses this method.

Another method is free float market capitalization, it excludes restricted shares by government, companies or employee stock options. The companies in the index are free floats based on their percentage of floats. Modified weighted capitalization also reduces the influence of large market capitalization companies. This method sets a limit on the weight of large stocks. NASDAQ 100 uses this method.

My evaluation period in this assessment is 1990: Q1 to 2020: Q1. The data sets are quarterly based (data tables are placed in the appendix).

However, I need to modify the simple data in order to use in Taylor rule equation.

**Output Gap:** The output gap estimation approach is to distinguish between the actual output and the pattern in which the output difference continues to revert. However, due to lack of availability of real-time data, I 'm following Linear Time Trend process.

I take the Real GDP Log and then regress it on a constant and linear time trend term with OLD estimator. Thus, the regression residuals multiplied by 100 percent, considered as output gap.

$$LGDP = \log(GDP)$$

$$Output\_Gap = 100 * (lgdp - hptrend)$$

**Inflation:** The Consumer Price Index (CPI) was among the most widely used inflation indicators. CPI measures mainly the weighted average market basket of changes to consumer products. It can also be a relatively good indicator of inflation, since there is real time limit for inflation.

To normalize CPI, I take CPI Log, then subtract from previous year to drive year-to-year change amounts. In addition, each country has its own inflation target according to its policies, but the inflation target is assumed to be 2% in developed countries, especially the United States and the United Kingdom. Finally, I calculate the CPI change by 100 and deduct (2 per cent) from the inflation target.

$$LCPI = \log(CPI)$$

$$Inflation\_Gap = 100 * (LCPI - LCPI(-4)) - 2\%$$

**Stock market:** There can be many stock market indicators to add in Taylor rule equation so as to examine the relationship between variables, such as Close price, P/E, P/B, EV/EBITDA, etc.

I considered Close Price on both FTSE100 and S&P500, which contains 100 and 500 large UK and US companies respectively. I assume closing price can be good indicator of the day and both Indexes are well diversified to consider as benchmark in the assessment.

So, after getting Log of each index, I subtract each period closing price from previous year, multiply by 100, to derive the % changed amount.

$$Asset = \log(s\&p500)$$

$$S = 100 * \big(Asset - Asset(-4)\big)$$

## Model

The original monetary policy rule of Taylor (1993) suggests that the central banks set a nominal short-term interest rate based on inflation rate adjustments and output gap. Therefore, forward-looking monetary policy guidelines linking the nominal interest rate to inflation and the output gap were more effective than the original backward-looking specification from Taylor (Orphanides 2003). A Taylor rule encompassing either current or forward-looking policy making is written as follows:

$$i(t) = \alpha + \beta(\pi(t) - \pi(t)^*) + \gamma(y(t) - y(t)^*) \qquad (1)$$

In this equation, i is the Short-Term Nominal interest rate, $\pi$ is the year-over-year inflation Rate, $\pi$ * is the Target Level of inflation (usually treated as a 2% constant in developed countries), y* is the percentage Deviation of output from its long-run trend (the output gap), $\alpha$ is the error term. Based on Clarida, Gali, and Gertler (1998), I believe that the central bank gradually adjusts the actual interest rate.

This strategy raises several challenges. First, real-time realized inflation values are not available to policymakers, so policymakers are pressured to be reluctant to change real-time interest rates. In addition, the realized inflation values are the "effect" of the Fed 's policy, not the "cause". Hence, they may not be suitable in the reaction to interest rates. For example, assume that both output gap and inflation are close to target rates, but the Greenbook forecast shows that the current policy situation is that I expect inflation to rise up to 4% next year, while the policy goal is to sustain it at 2%. *Ceteris paribus* said that rising inflation expectations will lead the Central Bank (Federal Reserve, Bank of England) to raise interest rates now, leading to actual inflation nearly 2 per cent next year. The value of price stabilization for Fed has risen since 1980 based on Figure La, as well as Greenbook 's inflation forecast has generally been above realized inflation when the inflation rate is above its target. This example illustrates the possible factor of bias when the Taylor rule calculation uses realized inflation of 2 per cent.

In the past couple of decades, the dramatic rise of share prices in the United States and United Kingdom has drawn people to stock markets as an economic proxy, especially through recession and economic shocks. It has contributed to a discussion about the potential association between equity prices and the form of policy rule implied by Taylor's rule. Some claim that fluctuations in asset markets may add more details to the Taylor rule. The probability of the relationship between asset price volatility and the actions of policy makers can be tested by adding new terms to the equation of Simple Taylor rule:

$$i(t) = \alpha + \beta(\pi(t) - \pi(t)^*) + \gamma(y(t) - y(t)^*) + \sum_{k=1}^{n} \delta(k)S(t-k) + \varepsilon(t) \quad (2)$$

Large and well diversified Stock Market indexes can be considered as Benchmark in both USA and UK, S&P500 and FTSE100 respectively. In the new term S(t-k) is year to year index change for stock market volatility, plus an error term, $\varepsilon$.

**Result**

USA Data Assessment

After explaining Taylor rule model, now I use data sets and apply Taylor rule equation (1) to assess the relationship between dependent variable (interest rate) and independent variables (inflation gap, output gap), then in the next stage add asset price volatility to the equation and examine the relationship.

Firstly, I plot the variables for USA data set as group to see any obvious relation:

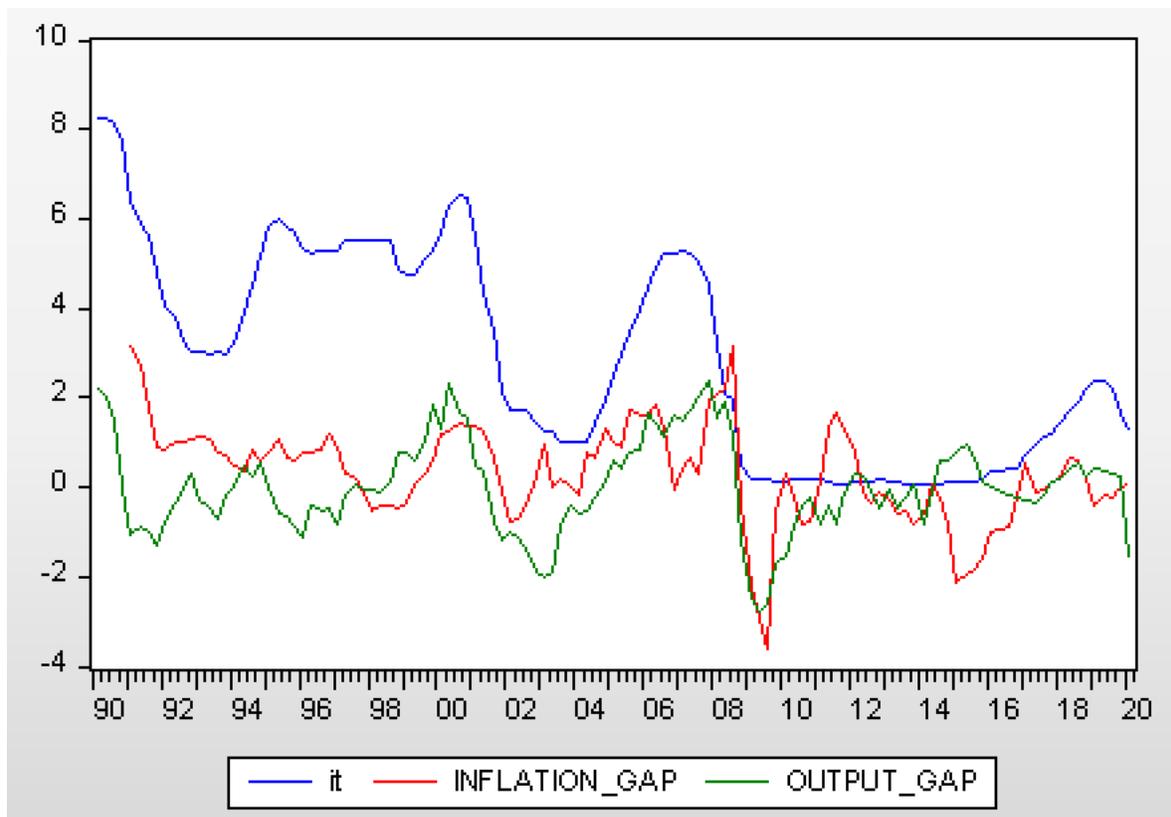

The graph illustrates that inflation gap and output gap have similar behavior in the U.S., as well as experiencing a significant decline in 1991, 2001, 2008, which can be explained by the presence of economic shocks such as recession in 1991 and 2001, and the financial crisis in 2008. Such shocks ultimately lead to a sharp decline in the inflation gap and output gap that prompted policymakers (FED) intervention to dramatically lower interest rates. The interest rate (IT) however decreased dramatically in 1991, 2001, 2008 and then stabilized.

Then I estimate the Taylor series equation without considering stock market index to find out the relationship between dependent and independent variables (Table 1):

| Dependent Variable: IT, Sample: 1991Q1 2020Q1 ||||
|---|---|---|---|
| Variable | Coefficient | t-Statistic | Prob |
| Inflation_Gap | 0.9063 | 5.4191 | 0 |
| OutPut_Gap | 0.454512 | 2.5263 | 0.0129 |
| C | 2.451161 | 13.6914 | 0 |
| R-squared | 0.3098 | Durbin-Watson stat | 0.1478 |
| Adjusted R-squared | 0.2977 | | |

The inflation coefficient is positive and significantly strong as expected which can explain the close association between inflation and interest rate.
An price rise of 1 percent corresponds to an interest rate rise of 0.91 per cent. In comparison, a unitary rise in the output gap increases the rate of interest by 0.45%.
In addition, I used R-squared to check the model, which calculates how close the data is to the fitted regression line, also known as the determination coefficient. I may note that the R2 shows that the model can explain about 31 per cent of the dependent variable variability. In fact, the F-statistic claims the difference between the mean of two population and is in favor of the model 's correctness, as its corresponding likelihood is smaller than 0.05. Lastly, the statistics for Durbin-Watson are around 0.05. This indicates that the specification can suffer from serial association, repetitive patterns.

Taylor (1993) suggest that the monetary policy should set both coefficients equal weight to 0.5. In order to test constraints on statistical parameters, I set a Wald test (Table 2):

| Wald Test | | |
|---|---|---|
| Null hypothesis: Coefficients are equally weighted | | |
| Test Statistic | Value | Probability |
| F-statistic | 3.1426 | 0.0469 |
| Chi-square | 6.2853 | 0.0432 |

In the Wald test the null hypothesis is equally weighted coefficients.
Given that Chi-square 's probability is smaller than 0.05, the Wald test is significant. So, in shaping monetary policy, I can reject the null hypothesis, and inflation and output gap weight differently.

Regarding the impression of the model, I want to see how it performs in predicting dependent variable variability. So, together with designed ones, I plot the actual values:

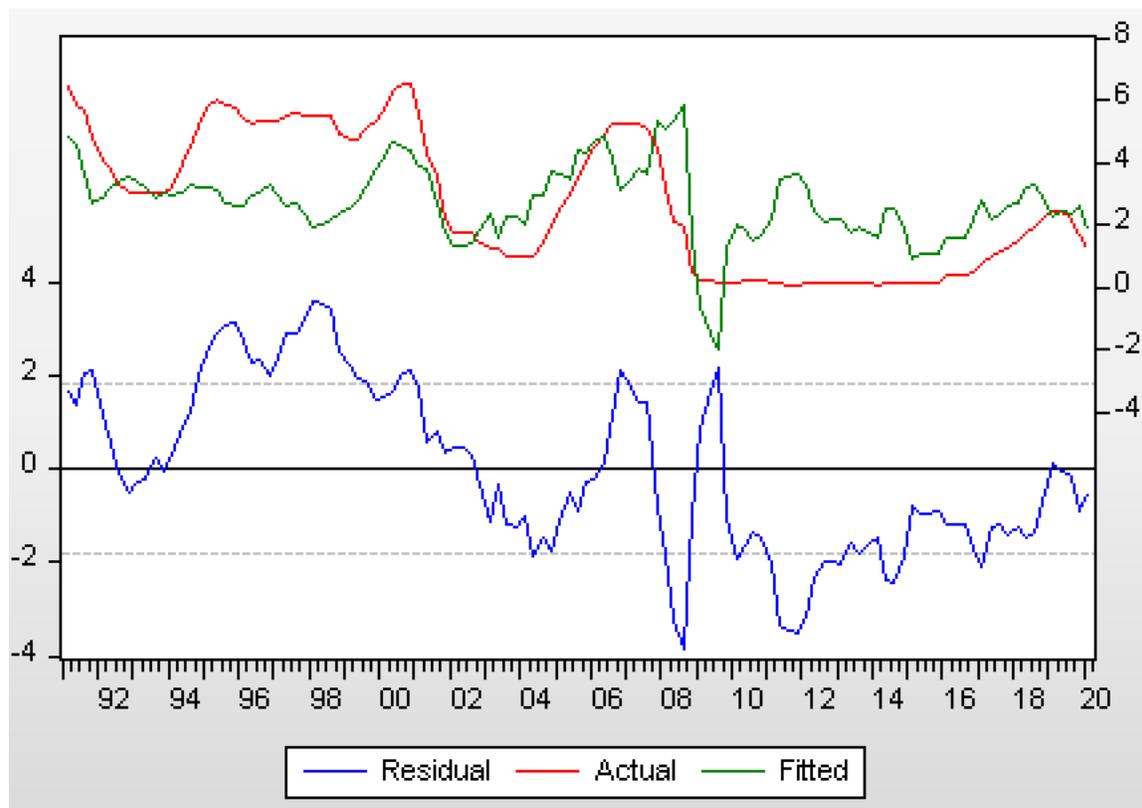

The model is the fitted value (green line). Fitted line underestimates the actual data (red line) from the beginning of 2003 and precisely matched the actual value from 3003 to 2009, while overestimating from 2006 onwards.

It may be perceived that the Taylor rule predicted lower interest rates before the break point and higher interest rates after the break.

I want to test the consistency of the link under investigation at the next point. More specifically it may be affected by the presence of certain shocks.

For instance, the recession at the beginning of 2003 and 2006 may had some consequences on the way in which the monetary authority shaped monetary policy. In order to disclose the presence of a structural break, I set up a Chow-Break test (Table 3):

| Chow Breakpoint Test: 2006Q1, Sample: 1991Q1 2020Q1 | | | |
|---|---|---|---|
| Null Hypothesis: No break at specified breakpoint | | | |
| F-statistic | 56.8077 | Prob. F(3.111) | 0.0000 |
| Wald Statistic | 170.4231 | Prob. Chi-Square(3) | 0.0000 |

or for 2006 (Table 4):

| Chow Breakpoint Test: 2006Q1, Sample: 1991Q1 2020Q1 | | | |
|---|---|---|---|
| Null Hypothesis: No break at specified breakpoint | | | |
| F-statistic | 30.3655 | Prob. F(3.111) | 0.0000 |
| Wald Statistic | 91.0965 | Prob. Chi-Square(3) | 0.0000 |

The null hypothesis for this test is No Break.

The null should be rejected, since the values are strongly significant. I would also assume that, in that period, some particular factors played a role in the way monetary policy was set.

According to the literature, by adding more information to the monetary authority, it is likely that asset price instability influences the monetary policy. In order to extract theory, stock market and interest-rate relationship can be investigated by adding additional term to the basic Taylor rule. To test the possibility, it is necessary to revisit equation to:

$$i(t) = \alpha + \beta(\pi(t) - \pi(t)^*) + \gamma(y(t) - y(t)^*) + \sum_{k=1}^{n} \delta(k) S(t-k) + \varepsilon(t)$$

In order to examine, I adopt a specific to general approach. I include lags, which particularly means delay, until they are not significant. So, start with one lag (Table 5):

| Dependent Variable: IT, Sample: 1991Q1 2020Q1 ||||
|---|---|---|---|
| Variable | Coefficient | t-Statistic | Prob |
| C | 2.3675 | 11.8876 | 0.0000 |
| Inflation_Gap | 0.8427 | 4.7943 | 0.0000 |
| OtPut_Gap | 0.4051 | 2.0023 | 0.0477 |
| S(-1) | 0.0117 | 0.9736 | 0.3323 |
| R-squared | 0.3035 | Durbin-Watson stat | 0.1417 |
| Adjusted R-squared | 0.2848 | | |

I look at the adjusted R2 instead of focusing on the R2, since the latter is sensitive only to the inclusion of meaningful regressors. I may note that this marginally decreased from 0.30 to 0.28, indicating that the model may not help describe the volatility of the dependent variable.

With respect to the coefficient associated with S(-1), I may mention that the t-statistic associate is equal to 0.97, with a p-value of 0.33. That means I do not have any credible evidence that there is a strong relationship between asset price and interest rate. It is also sufficient to argue that uncertainty in asset prices (S&P500) is not adding more knowledge into the model.

Given the result, I do not re-estimate my model by including the second lag of S. Therefore, I can conclude that no lag should be included in the model.

A source of misspecification may come from the fact that the model does not satisfy the main assumptions of classical regression model. Specifically, it may suffer from heteroskedasticity and serial correlation of the residuals and at a small extent from normality of the residuals. So, I carry out some tests to check the model.

The first test is normality test:

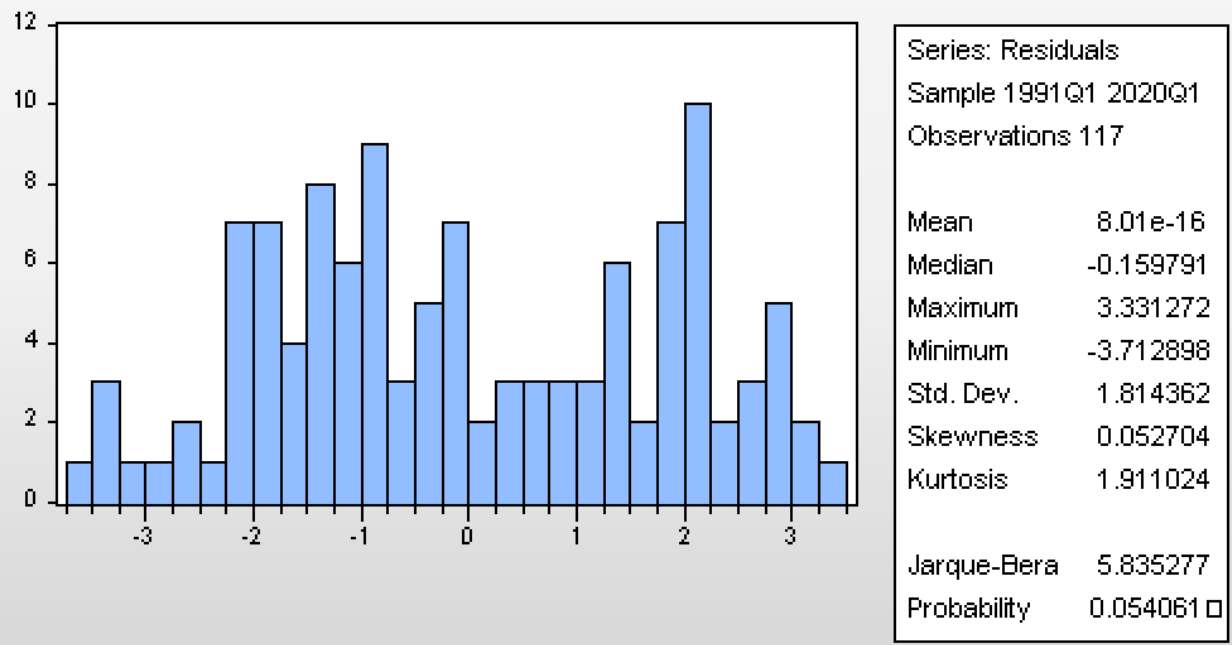

In normality test Jaraque Bera test the goodness of fit test through matching skewness and kurtosis with a normal distribution, and also it has the null of normally distribution in the residuals.

Jaraque bera test reject the hypothesis of normality in the distribution of the residuals. Actually, I could reach the same result by looking at the graph reported along with statistic.

The second test is White test for heteroskedasticity (Table 6):

| Heteroskedasticity Test: White | | | |
|---|---|---|---|
| Null hypothesis: Homoskedasticity | | | |
| Fstatistic | 4.1786 | Prob. F(9,107) | 0.0001 |
| Obs R-squared | 30.4280 | Prob Chi-Square(9) | 0.0004 |

Homoskedasticity is where the variance of errors is constant. Homoskedasticity is the null hypothesis, and I can clearly reject the null hypothesis and thus the residuals are heteroskedastic. The volatility level cannot therefore be predicted for any period.

The third test is LM test for serial correlation (Table 7):

As I did not include any lags, I should put 0 to the number of lags.

| Breusch-Godfrey Serial Correlation LM Test: | | | |
|---|---|---|---|
| Null hypothesis: No serial correlation at up to 1 lag | | | |
| F-statistic | 650.9373 | Prob. F(1,112) | 0.0000 |
| Obs*R-squared | 99.8242 | Prob Chi-square(1) | 0.0000 |

I can strongly say I have enough evidence to refute the null hypothesis, No Serial Correlation. So, serial correlation problem exist which means the variables are not independent from one another.

I may conclude that the model suffers from both serial correlation and heteroskedasticity, mean the future observations are affected by past values and the volatility cannot be predicted!

I may overcome this problem by re-estimating the model using the Newey-West option to obtain robust standard errors by estimating covariance matrix of variables (Table 8):

| Dependent variable: IT ,Sample: 1991Q1 2020Q1 | | | |
|---|---|---|---|
| HAC standard errors & covariance (Bartlett kernel, Newey-West fixed bandwith=5.0000 | | | |
| Variable | Coefficient | t-Statistic | Prob |
| C | 2.3638 | 7.6767 | 0.0000 |
| Inflation_Gap | 0.8911 | 2.9363 | 0.0040 |
| OutPut_Gap | 0.3979 | 1.2832 | 0.2020 |
| S | 0.0120 | 0.5879 | 0.5577 |

I may conclude that after re-estimating the model by using robust standard errors, just inflation gap variable enters the model significantly.

Asset price volatility affects output gap. Therefore, there exists a problem of endogeneity, explanatory variable is correlated with the error term, that I need to address. I achieve so by using a GMM estimator to apply an IV method, which blends independent variables with population-time information to determine dependent parameters (Table 9):

| Dependent variable: IT ,Sample: 1991Q3 2020Q1 | | | |
|---|---|---|---|
| Estimation Weighting Matrix: HAC (Bartlett Kernel, New-West fixed bandwidth=5.0000) | | | |
| Instrument Specification: Inflation_Gap(-1) Inflation_Gap(-2) OutPut_Gap(-1) OutPut_Gap(-2) | | | |
| Variable | Coefficient | t-Statistic | Prob |
| C | 2.8076 | 6.2563 | 0.0000 |
| Inflation_Gap | 0.8071 | 1.8996 | 0.0601 |
| OutPut_Gap | 0.9315 | 2.3857 | 0.0187 |
| S | -0.0526 | -1.7846 | 0.0770 |
| J-statistic | 3.6830 | Prob (J-statistic) | 0.055 |

The value associated with J-statistic is the value of the GMM objective function evaluated using an efficient GMM estimator, and so it supports the choice of the instruments, as the probability is higher than 0.05.

The S-related coefficient is therefore negative, and not statistically significant. It may be evidence that the volatility of asset prices is not influencing the interest rate.

Therefore, the key result is that adding equity prices to Taylor's rule equation does not strengthen the model to help justify the Fed Fund's rate decision, and by generating indeterminacy in reasonable assumptions it would hurt economic efficiency.

## UK Data Assessment

In this stage I implement the same procedure for the UK data sets.
after calculating each variable (output gap, inflation gap) for Taylor series equation, I plot the variables of interest between 1990Q1 and 2020Q1:

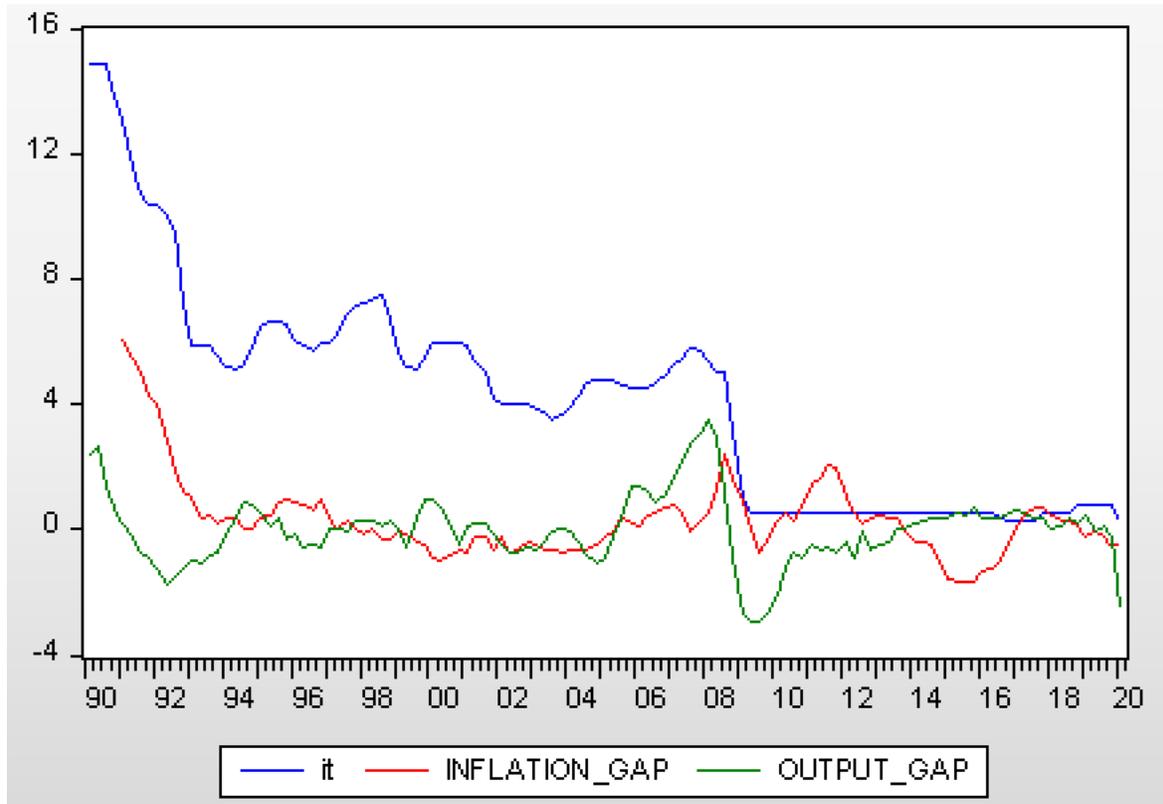

Inflation gap and output gap display similar behavior to that of the USA, and they have suffered a substantial downturn in 2008, which can be explained by the existence of a financial crisis. In fact, the financial downturn has led to a dramatic fall in inflation and has culminated in policymakers taking steps to dramatically lower the interest rate. Nevertheless, the interest rate (IT) plummeted sharply in 2008 and stabilized afterwards.

Then estimating the Taylor Series equation without considering Stock Market Index can reveal the relationship between dependent and independent variables (Table 10):

| Dependent Variable: IT, Sample: 1991Q1 2020Q1 ||||
|---|---|---|---|
| Variable | Coefficient | t-Statistic | Prob |
| Inflation_Gap (%) | 1.2564 | 6.7208 | 0 |
| OutPut_Gap (%) | 0.514578 | 2.270698 | 0.025 |
| C | 3.4587 | 14.123 | 0 |
| R-squared (%) | 0.2927 | Durbin-Watson stat | 0.0489 |
| Adjusted R-squared (%) | 0.2804 | | |

As expected, the coefficient to inflation is positive and significantly strong which can explain the strong relationship between inflation and interest rate.
An inflation increase of 1%, results in an interest rate increase of 1.26%. In fact, a unitary rise in the output gap increases the interest rate by 0.51%.

Additionally, I used R-squared to test the formula, which calculates how close the data is to the fitted regression line, also known as the determination coefficient. R2 shows that the model will describe the volatility of the dependent variable marginally lower (29 per cent) relative to the USA model (31 per cent).
In general, the F-statistic, the differential between mean of two population, is in favor of model validity, as the likelihood is smaller than 0.05. Eventually the figures for Durbin-Watson are about 0.05. This may be a sign of serial correlation, repeated patterns.

According to the simple Taylor (1993), monetary policy should set both coefficients equal weight to 0.5. In order to test constraints on statistical parameters a Wald test can be set (Table 11):

| Wald Test |||
|---|---|---|
| Null hypothesis: Coefficients are equally weighted |||
| Test Statistic | Value | Probability |
| F-statistic | 97.68412 | 0 |
| Chi-square | 195.3682 | 0 |

In the Wald test the null hypothesis is equally weighted coefficients.
Given that Chi-square 's probability is smaller than 0.05, the Wald test is significant. I can therefore reject the null hypothesis, coefficients are weighted equally, and thus inflation and output gap weight differently in monetary policy shaping.

Regarding the model impression, I want to see how it performs in predicting the variability of the interest rate. So, I plot the actual values along with fitted ones:

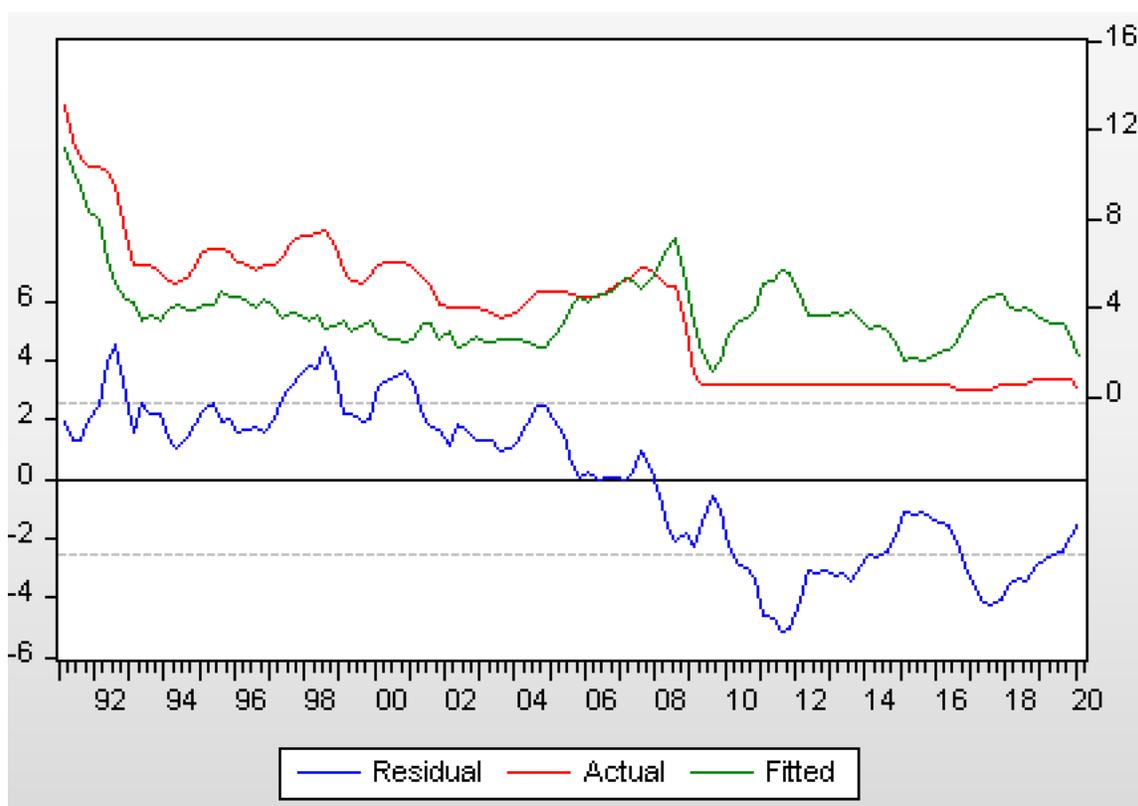

The model is the fitted value (green line). Unlike the USA model that fit the actual data closely, Fitted line is underestimated the actual data (red line) from the beginning, while it overestimates from 2006 onwards in UK.
It can be assumed, Taylor rule predicted lower interest rate before break point and higher interest rate after break point.

As I did for USA, I want to check the stability of the relationship under investigation. Particularly, the presence of some shocks may affect the way that policy makers set the interest rate.

For instance, the financial crisis at the beginning of 2006 may had some consequences on the way in which the monetary authority shaped monetary policy. Chow-Break test can disclose the presence of structural break (Table 12):

| Chow Breakpoint Test: 2006Q1, Sample: 1991Q1 2020Q1 | | | |
|---|---|---|---|
| Null Hypothesis: No break at specified breakpoint | | | |
| F-statistic | 164.4359 | Prob. F(3.111) | 0.0000 |
| Wald Statistic | 493.3077 | Prob. Chi-Square(3) | 0.0000 |

The null hypothesis for this test is No Break.

The null is rejected due to significant values. I may also assume that, in that period, certain particular factors played a part in the way monetary policy was set.

Again, by adding more information to the monetary authority, it is possible that uncertainty in asset prices impacts monetary policy. I revise the equation to check the possibility:

$$i(t) = \alpha + \beta(\pi(t) - \pi(t)^*) + \gamma(y(t) - y(t)^*) + \sum_{k=1}^{n} \delta(k)S(t-k) + \varepsilon(t)$$

Specific to general approach indicates to include lags, until they are not significant. So, start with one lag (Table 13):

| Dependent Variable: IT, Sample: 1991Q1 2020Q1 | | | |
|---|---|---|---|
| Variable | Coefficient | t-Statistic | Prob |
| C | 3.3671 | 13.3325 | 0.0000 |
| Inflation_Gap | 1.1247 | 5.3573 | 0.0000 |
| OtPut_Gap | 0.3883 | 1.6216 | 0.1077 |
| S(-1) | 0.0254 | 1.4373 | 0.1534 |
| R-squared | 0.2491 | Durbin-Watson stat | 0.0509 |
| Adjusted R-squared | 0.2290 | | |

Like in the USA section, I look the adjusted $R^2$. I may note that it decreased from 0.28 to 0.23, thus suggesting that the model is not explaining better the variability of the dependent variable through adding lags.

Since the coefficient associated to S(-1) is concerned, the associate t-statistic is equal to 1.44, with a p-value of 0.15. That means I do not have any meaningful evidence that there is a strong relationship between asset price and interest rate. This is also adequate to assert that the uncertainty of the asset price (FTSE100) does not carry further knowledge into the model, such as the USA test.

Given the result, I do not re-estimate the model by including the second lag of S. Hence, I may conclude that no lag should be included in the model.

This may be concluded that a cause of mis-specification arises from the fact that the model does not follow the core assumptions of the concept of classical regression. Specifically, it may suffer from residual heteroskedasticity and serial correlation, and to a small extent from residual normality. So, I do some testing again to check the model.

Normality test:

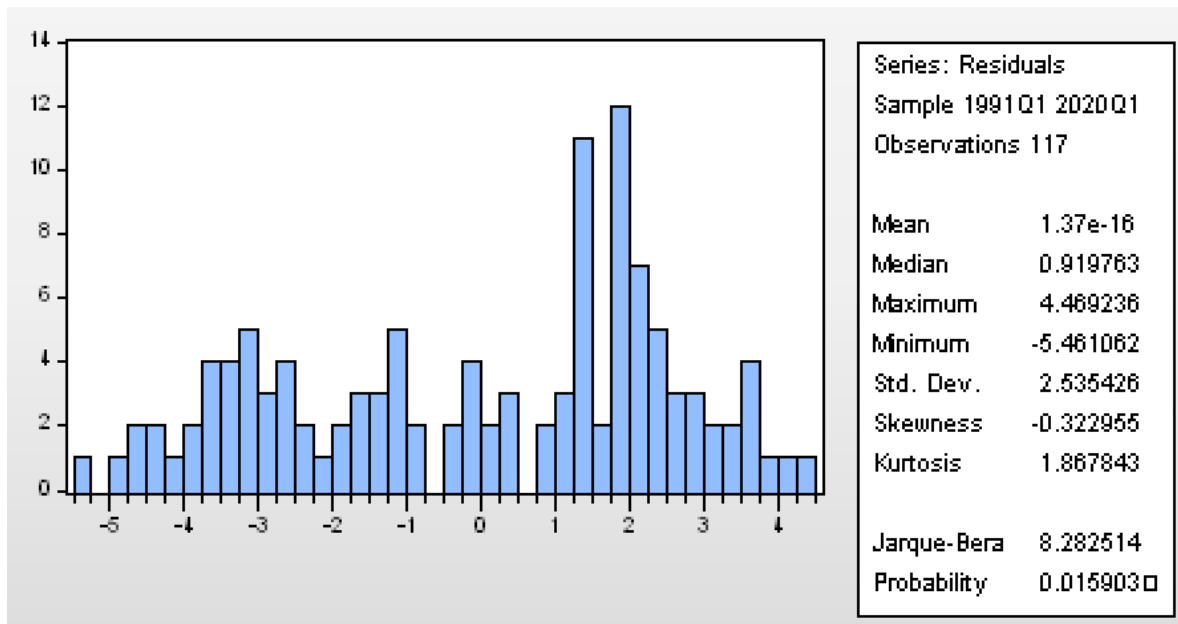

As Jaraque Bera test is explained in USA section, Jaraque bera test reject the hypothesis of normality in the distribution of the residuals because of 0.016 probability. Obviously, the result could be seen by the graph reported along with statistic.

The second test is White test for heteroskedasticity (Table 14):

| Heteroskedasticity Test: White |||| 
|---|---|---|---|
| Null hypothesis: Homoskedasticity ||||
| Fstatistic | 4.2235 | Prob. F(9,107) | 0.0001 |
| Obs R-squared | 30.6689 | Prob Chi-Square(9) | 0.0003 |

The null hypothesis is Homoskedasticity, variance of errors is constant, so I can clearly reject the null hypothesis (0.00 Probability), and so the residuals are heteroskedastic. Thus, the level of volatility cannot be predicted over any period.

The last test is LM test for serial correlation (Table 15):

As I did not include any lags, I should put 0 to the number of lags like USA model:

| Breusch-Godfrey Serial Correlation LM Test: | | | |
|---|---|---|---|
| Null hypothesis: No serial correlation at up to 1 lag | | | |
| F-statistic | 1702.7310 | Prob. F(1,112) | 0.0000 |
| Obs*R-squared | 109.7791 | Prob Chi-square(1) | 0.0000 |

I may firmly state that I have ample proof to refute the null hypothesis, No Serial Correlation. There is also an issue with serial correlation which implies that the variables are not independent from each other.

I may infer that the model suffers from both serial correlation and heteroskedasticity, indicating that past values influence future findings and the volatility cannot be anticipated!

I may resolve this issue by re-estimating the model using the Newey-West option to get robust standard errors by estimating the covariance matrix of variables (Table 16):

| Dependent variable: IT ,Sample: 1991Q1 2020Q1 | | | |
|---|---|---|---|
| HAC standard errors & covariance (Bartlett kernel, Newey-West fixed bandwith=5.0000 | | | |
| Variable | Coefficient | t-Statistic | Prob |
| C | 3.3837 | 6.8462 | 0.0000 |
| Inflation_Gap | 1.2281 | 3.9292 | 0.0002 |
| OutPut_Gap | 0.4645 | 1.7165 | 0.0888 |
| S | 0.0198 | 0.7831 | 0.4352 |

I may conclude that, after re-estimating the model with the use of robust standard errors, inflation gap and output gap variables significantly join the formula, unlike the USA, where only inflation gap substantially joins.

Volatility of asset prices influences output gap. And so, there is an endogeneity issue that needs to be discussed. I do so by using a GMM estimator to execute an IV method, which blends independent variables with data in population moment conditions to determine dependent parameters (Table 17):

| Dependent variable: IT, Sample: 1991Q3 2020Q1 | | | |
|---|---|---|---|
| Estimation Weighting Matrix: HAC (Bartlett Kernel, New-West fixed bandwidth=5.0000) | | | |
| Instrument Specification: Inflation_Gap(-1) Inflation_Gap(-2) OutPut_Gap(-1) OutPut_Gap(-2) | | | |
| Variable | Coefficient | t-Statistic | Prob |
| C | 3.6271 | 6.6122 | 0.0000 |
| Inflation_Gap | 1.1325 | 2.4125 | 0.0175 |
| OutPut_Gap | 0.5798 | 1.5437 | 0.1255 |
| S | -0.0165 | -0.2898 | 0.7725 |
| J-statistic | 2.3971 | Prob (J-statistic) | 0.1216 |

The value associated with J-statistic is the value of the GMM objective function evaluated using an efficient GMM estimator, and so it supports the choice of the instruments (probability higher than 0.05) like USA.

Besides, asset price volatility does not affect the interest rate because the coefficient associated with S is negative and not significant.

Therefore, I achieve the same result that adding equity prices to the Taylor rule equation will not boost the model to better explain the policy rate decision, and even causing indeterminacy of rational expectations could harm economic performance.

**Conclusion**

The Taylor rule is a descriptive model of monetary policy on the central bank. Simplicity and transparency are the Taylor rule's most important feature. Also, it can be used not only to test the decisions made by monetary authorities, but also to minimize general population policy decisions confusion because of its simplicity. While interpretation of monetary policy has become a helpful method, it should not blindly obey simple policy rule. The number of variables in Taylor rule is limited and it does not prevent further policy errors. In addition, additional macroeconomic indicators could provide more information to optimize the model. For example, solvency and liquidity problem may arise by considerable changes in asset prices. So, considering asset price volatility can optimize the model.

This paper looks at the relationship between Taylor rule equation variables and asset price fluctuations in the USA and UK from Q1:1990 to Q1:2020. Economic uncertainty provides policy-makers with challenges. Given the statement, I begin with driving output gap, inflation, and stock market volatility, then calculate the equation of Taylor's rule to find out the coefficients and the robustness of the model. Since both output gap and inflation coefficients are significant, there is a strong relationship between dependent and independent variables such that independent variables with high precision can describe the interest rate adjustments (R-squared 30% USA, 28% UK). Then I check the stability of model in critical points, shocking period, through Chow-Break test. In the next stage, I add additional term to Taylor rule equation in order to assess the relationship between asset price, stock market index is used as diversified benchmark, and interest rate. First, I check the robustness of the model. As the asset price does not improve the model (R-squared 28% USA, 23% UK), different tests and procedures use to find the source of misspecification, such as Normality test, White test, re-estimation the model by Newey-West, and using GMM estimator. Ultimately, I come up with the conclusion that asset price volatility does not affect the interest rate and so including asset price would not optimize the model in both countries.

According to the literature analysis, there is no definitive response as to whether the central bank should or should not take on account assets. Some economists believe central bank should not pay attention to asset price, because the interest rate would stabilize the asset prices itself, also, the reason behind shift in asset prices is not clear, so stabilizing asset prices may lead to unpredictable volatility. The rest, however, are of the opinion that shock cycle is different for non-fundamental reasons. So, considering asset prices can boost the efficiency. Lastly, the economic data suggests that the policy rate is set primarily in reaction to difference in inflation and output gap.

**Appendix**

## Data Table (1)

### Variable Definition and Sample Statistics, USA

Federal Reserve Economic Data, Link: https://fred.stlouisfed.org
Frequency: Quarterly
Observation Data: 1990 Q1- 2020 Q1
Real GDP: Billions of Chained 2012 Dollars Quarterly, Seasonally Adjusted Annual Rate
CPI Index: CPI Index for All Urban Consumers: All Items in U.S. City Average, Index 1982-1984=100, Seasonally Adjusted
Interest Rate (IT): Effective Federal Funds Rate, Percent, Not Seasonally Adjusted
S&P500: Close price, 500 large US corporations, Quarterly adjusted

| Observation Date | Real GDP | CPI Index | Interest Rate (IT) | S&P500 Index |
|---|---|---|---|---|
| 1/1/90 | 9358.289 | 128.033 | 8.248 | 353.400 |
| 4/1/90 | 9392.251 | 129.300 | 8.239 | 339.940 |
| 7/1/90 | 9398.499 | 131.533 | 8.160 | 358.020 |
| 10/1/90 | 9312.937 | 133.767 | 7.743 | 306.050 |
| 1/1/91 | 9269.367 | 134.767 | 6.430 | 330.220 |
| 4/1/91 | 9341.642 | 135.567 | 5.864 | 375.220 |
| 7/1/91 | 9388.845 | 136.600 | 5.645 | 371.160 |
| 10/1/91 | 9421.565 | 137.733 | 4.818 | 387.860 |
| 1/1/92 | 9534.346 | 138.667 | 4.024 | 417.090 |
| 4/1/92 | 9637.732 | 139.733 | 3.774 | 403.690 |
| 7/1/92 | 9732.979 | 140.800 | 3.259 | 408.140 |
| 10/1/92 | 9834.510 | 142.033 | 3.035 | 417.800 |
| 1/1/93 | 9850.973 | 143.067 | 3.042 | 435.710 |
| 4/1/93 | 9908.347 | 144.100 | 2.997 | 451.670 |
| 7/1/93 | 9955.641 | 144.767 | 3.058 | 450.530 |
| 10/1/93 | 10091.049 | 145.967 | 2.988 | 458.930 |
| 1/1/94 | 10188.954 | 146.700 | 3.209 | 466.450 |
| 4/1/94 | 10327.019 | 147.533 | 3.938 | 445.770 |
| 7/1/94 | 10387.382 | 148.900 | 4.485 | 444.270 |
| 10/1/94 | 10506.372 | 149.767 | 5.168 | 462.690 |
| 1/1/95 | 10543.644 | 150.867 | 5.803 | 459.270 |
| 4/1/95 | 10575.100 | 152.100 | 6.019 | 500.710 |
| 7/1/95 | 10665.060 | 152.867 | 5.797 | 544.750 |
| 10/1/95 | 10737.478 | 153.700 | 5.719 | 584.410 |
| 1/1/96 | 10817.896 | 155.067 | 5.371 | 615.930 |
| 4/1/96 | 10998.322 | 156.400 | 5.244 | 645.500 |
| 7/1/96 | 11096.976 | 157.300 | 5.306 | 670.630 |
| 10/1/96 | 11212.205 | 158.667 | 5.281 | 687.310 |
| 1/1/97 | 11284.587 | 159.633 | 5.278 | 740.740 |
| 4/1/97 | 11472.137 | 160.000 | 5.522 | 757.120 |
| 7/1/97 | 11615.636 | 160.800 | 5.535 | 885.140 |
| 10/1/97 | 11715.393 | 161.667 | 5.507 | 947.280 |
| 1/1/98 | 11832.486 | 162.000 | 5.519 | 970.430 |
| 4/1/98 | 11942.032 | 162.533 | 5.497 | 1101.750 |
| 7/1/98 | 12091.614 | 163.367 | 5.532 | 1133.840 |
| 10/1/98 | 12287.000 | 164.133 | 4.861 | 1017.010 |
| 1/1/99 | 12403.293 | 164.733 | 4.735 | 1229.230 |
| 4/1/99 | 12498.694 | 165.967 | 4.748 | 1286.370 |
| 7/1/99 | 12662.385 | 167.200 | 5.096 | 1372.710 |
| 10/1/99 | 12877.593 | 168.433 | 5.304 | 1282.710 |
| 1/1/00 | 12924.179 | 170.100 | 5.678 | 1469.250 |
| 4/1/00 | 13160.842 | 171.433 | 6.272 | 1498.580 |
| 7/1/00 | 13178.419 | 173.000 | 6.519 | 1454.600 |
| 10/1/00 | 13260.506 | 174.233 | 6.475 | 1436.510 |
| 1/1/01 | 13222.690 | 175.900 | 5.597 | 1320.280 |
| 4/1/01 | 13299.984 | 177.133 | 4.327 | 1160.330 |
| 7/1/01 | 13244.784 | 177.633 | 3.502 | 1224.420 |
| 10/1/01 | 13280.859 | 177.500 | 2.130 | 1040.940 |
| 1/1/02 | 13397.002 | 178.067 | 1.733 | 1148.080 |
| 4/1/02 | 13478.152 | 179.467 | 1.752 | 1147.390 |
| 7/1/02 | 13538.072 | 180.433 | 1.741 | 989.810 |
| 10/1/02 | 13559.032 | 181.500 | 1.444 | 815.280 |
| 1/1/03 | 13634.253 | 183.367 | 1.250 | 879.820 |
| 4/1/03 | 13751.543 | 183.067 | 1.247 | 848.180 |
| 7/1/03 | 13985.073 | 184.433 | 1.017 | 974.500 |
| 10/1/03 | 14145.645 | 185.133 | 0.997 | 995.970 |
| 1/1/04 | 14221.147 | 186.700 | 1.002 | 1111.920 |
| 4/1/04 | 14329.523 | 188.167 | 1.011 | 1126.210 |
| 7/1/04 | 14464.984 | 189.367 | 1.431 | 1140.840 |
| 10/1/04 | 14609.876 | 191.400 | 1.950 | 1114.580 |
| 1/1/05 | 14771.602 | 192.367 | 2.469 | 1211.920 |
| 4/1/05 | 14839.782 | 193.667 | 2.942 | 1180.590 |
| 7/1/05 | 14972.054 | 196.600 | 3.460 | 1191.330 |
| 10/1/05 | 15066.597 | 198.433 | 3.978 | 1228.810 |
| 1/1/06 | 15267.026 | 199.467 | 4.454 | 1248.290 |
| 4/1/06 | 15302.705 | 201.267 | 4.908 | 1294.830 |
| 7/1/06 | 15326.368 | 203.167 | 5.245 | 1270.200 |
| 10/1/06 | 15456.928 | 202.333 | 5.243 | 1335.850 |
| 1/1/07 | 15493.328 | 204.317 | 5.255 | 1418.300 |
| 4/1/07 | 15582.085 | 206.631 | 5.252 | 1420.860 |
| 7/1/07 | 15666.738 | 207.939 | 5.074 | 1503.350 |
| 10/1/07 | 15761.967 | 210.490 | 4.496 | 1526.750 |
| 1/1/08 | 15671.383 | 212.770 | 3.181 | 1468.360 |
| 4/1/08 | 15752.308 | 215.538 | 2.085 | 1322.700 |
| 7/1/08 | 15667.032 | 218.861 | 1.941 | 1280.000 |
| 10/1/08 | 15328.027 | 213.849 | 0.505 | 1166.360 |
| 1/1/09 | 15155.940 | 212.378 | 0.184 | 903.250 |
| 4/1/09 | 15134.117 | 213.507 | 0.178 | 797.870 |
| 7/1/09 | 15189.222 | 215.344 | 0.154 | 919.320 |
| 10/1/09 | 15356.058 | 217.030 | 0.118 | 1057.080 |
| 1/1/10 | 15415.145 | 217.374 | 0.134 | 1115.100 |
| 4/1/10 | 15557.277 | 217.297 | 0.192 | 1169.430 |
| 7/1/10 | 15671.967 | 217.934 | 0.189 | 1030.710 |
| 10/1/10 | 15750.625 | 219.699 | 0.190 | 1141.200 |
| 1/1/11 | 15712.754 | 222.044 | 0.155 | 1257.640 |
| 4/1/11 | 15825.096 | 224.568 | 0.095 | 1325.830 |
| 7/1/11 | 15820.700 | 226.033 | 0.084 | 1320.640 |
| 10/1/11 | 16004.107 | 227.047 | 0.074 | 1131.420 |
| 1/1/12 | 16129.418 | 228.326 | 0.104 | 1257.610 |
| 4/1/12 | 16198.807 | 228.808 | 0.152 | 1408.470 |
| 7/1/12 | 16220.667 | 229.841 | 0.144 | 1362.160 |
| 10/1/12 | 16239.138 | 231.369 | 0.161 | 1440.670 |
| 1/1/13 | 16382.964 | 232.299 | 0.144 | 1426.190 |
| 4/1/13 | 16403.180 | 232.045 | 0.116 | 1569.190 |
| 7/1/13 | 16531.685 | 233.300 | 0.085 | 1606.280 |
| 10/1/13 | 16663.649 | 234.163 | 0.086 | 1681.550 |
| 1/1/14 | 16616.540 | 235.621 | 0.072 | 1848.360 |
| 4/1/14 | 16841.475 | 236.872 | 0.091 | 1872.340 |
| 7/1/14 | 17047.098 | 237.478 | 0.089 | 1960.230 |
| 10/1/14 | 17143.038 | 236.888 | 0.101 | 1972.290 |
| 1/1/15 | 17277.580 | 235.355 | 0.113 | 2058.900 |
| 4/1/15 | 17405.669 | 236.960 | 0.126 | 2067.890 |
| 7/1/15 | 17463.222 | 237.855 | 0.135 | 2063.110 |
| 10/1/15 | 17468.902 | 237.837 | 0.161 | 1920.030 |
| 1/1/16 | 17556.839 | 237.777 | 0.360 | 2043.940 |
| 4/1/16 | 17639.417 | 239.473 | 0.369 | 2059.740 |
| 7/1/16 | 17735.074 | 240.591 | 0.395 | 2098.860 |
| 10/1/16 | 17824.231 | 242.115 | 0.448 | 2168.270 |
| 1/1/17 | 17925.256 | 243.822 | 0.699 | 2238.830 |
| 4/1/17 | 18021.048 | 244.054 | 0.947 | 2362.720 |
| 7/1/17 | 18163.558 | 245.359 | 1.154 | 2423.410 |
| 10/1/17 | 18322.464 | 247.250 | 1.205 | 2519.360 |
| 1/1/18 | 18438.254 | 249.235 | 1.447 | 2673.610 |
| 4/1/18 | 18598.135 | 250.591 | 1.737 | 2640.870 |
| 7/1/18 | 18732.720 | 251.883 | 1.926 | 2718.370 |
| 10/1/18 | 18783.548 | 252.697 | 2.220 | 2913.980 |
| 1/1/19 | 18927.281 | 253.275 | 2.402 | 2506.850 |
| 4/1/19 | 19021.860 | 255.171 | 2.397 | 2834.400 |
| 7/1/19 | 19121.112 | 256.325 | 2.192 | 2941.760 |
| 10/1/19 | 19221.970 | 257.832 | 1.646 | 2976.740 |
| 1/1/20 | 18974.702 | 258.608 | 1.255 | 3230.780 |

**Reference**: Federal Reserve Bank of St. Louis, Link: https://fred.stlouisfed.org

## Data Table (2)

### Variable Definition and Statistics, UK

Federal Reserve Economic Data, Link: https://fred.stlouisfed.org
Frequency: Quarterly
Observation Data: 1990 Q1- 2020 Q1
Real GDP: Millions of Chained 2010 National Currency, Seasonally Adjusted
CPI Index: CPI Index, Index 2015=100, Not Seasonally Adjusted
Interest Rate (IT): Bank of England Policy Rate in the United Kingdom, Percent per Annum, Not Seasonally Adjusted
FTSE100: Close price, 100 large UK corporations, Quarterly adjusted

| Observation Date | Real GDP | CPI Index | Interest rate (IT) | FTSE100 Index |
|---|---|---|---|---|
| 1990-01-01 | 266386.8 | 52.8000 | 14.88 | 2422.7 |
| 1990-04-01 | 267755.2 | 54.8667 | 14.88 | 2247.9 |
| 1990-07-01 | 264959.8 | 55.7000 | 14.88 | 2374.7 |
| 1990-10-01 | 264048.2 | 56.8667 | 13.88 | 1990.3 |
| 1991-01-01 | 263277.6 | 57.2333 | 13.05 | 2143.4 |
| 1991-04-01 | 262945.7 | 59.1333 | 11.55 | 2456.6 |
| 1991-07-01 | 262371.5 | 59.7667 | 10.71 | 2414.7 |
| 1991-10-01 | 262827.3 | 60.5333 | 10.38 | 2621.7 |
| 1992-01-01 | 262846.3 | 60.8000 | 10.38 | 2493.1 |
| 1992-04-01 | 262536.1 | 62.0667 | 10.05 | 2440.1 |
| 1992-07-01 | 264208.2 | 62.1333 | 9.55 | 2521.2 |
| 1992-10-01 | 266048.6 | 62.5333 | 7.21 | 2553 |
| 1993-01-01 | 267973.1 | 62.7000 | 5.88 | 2846.5 |
| 1993-04-01 | 269301.6 | 63.5333 | 5.88 | 2878.7 |
| 1993-07-01 | 271423.3 | 63.7000 | 5.88 | 2900 |
| 1993-10-01 | 273224.8 | 63.9333 | 5.55 | 3037.5 |
| 1994-01-01 | 276437.1 | 64.2000 | 5.21 | 3418.4 |
| 1994-04-01 | 279663.0 | 65.0333 | 5.13 | 3086.4 |
| 1994-07-01 | 282859.0 | 65.0000 | 5.30 | 2919.2 |
| 1994-10-01 | 284574.6 | 65.2667 | 5.80 | 3026.3 |
| 1995-01-01 | 285782.8 | 65.7333 | 6.46 | 3065.5 |
| 1995-04-01 | 286839.1 | 66.6333 | 6.63 | 3137.9 |
| 1995-07-01 | 289489.8 | 66.9000 | 6.63 | 3314.6 |
| 1995-10-01 | 289862.4 | 67.2333 | 6.55 | 3508.2 |
| 1996-01-01 | 292524.0 | 67.6667 | 6.07 | 3689.3 |
| 1996-04-01 | 293565.8 | 68.5333 | 5.86 | 3699.7 |
| 1996-07-01 | 296152.3 | 68.6667 | 5.69 | 3711 |
| 1996-10-01 | 298434.0 | 69.2333 | 5.94 | 3953.7 |
| 1997-01-01 | 302554.3 | 69.3000 | 5.94 | 4118.5 |
| 1997-04-01 | 305166.1 | 69.9333 | 6.23 | 4312.9 |
| 1997-07-01 | 307409.0 | 70.2667 | 6.92 | 4604.6 |
| 1997-10-01 | 311059.9 | 70.6333 | 7.17 | 5244.2 |
| 1998-01-01 | 313759.4 | 70.6000 | 7.25 | 5135.5 |
| 1998-04-01 | 316527.7 | 71.3667 | 7.33 | 5932.2 |
| 1998-07-01 | 318689.1 | 71.4333 | 7.50 | 5832.5 |
| 1998-10-01 | 321906.0 | 71.8333 | 6.75 | 5064.4 |
| 1999-01-01 | 323936.3 | 72.0000 | 5.67 | 5882.6 |
| 1999-04-01 | 324604.6 | 72.7000 | 5.17 | 6295.3 |
| 1999-07-01 | 330573.4 | 72.6000 | 5.08 | 6318.5 |
| 1999-10-01 | 335336.7 | 72.9333 | 5.42 | 6029.8 |
| 2000-01-01 | 337891.6 | 72.8000 | 5.92 | 6930.2 |
| 2000-04-01 | 339782.6 | 73.4667 | 6.00 | 6540.2 |
| 2000-07-01 | 340711.4 | 73.4667 | 6.00 | 6312.7 |
| 2000-10-01 | 341247.7 | 73.9333 | 6.00 | 6294.2 |
| 2001-01-01 | 346129.4 | 73.7000 | 5.83 | 6222.5 |
| 2001-04-01 | 349025.2 | 74.7667 | 5.33 | 5633.7 |
| 2001-07-01 | 351806.2 | 74.7667 | 5.00 | 5642.5 |
| 2001-10-01 | 353112.1 | 74.9333 | 4.17 | 4903.4 |
| 2002-01-01 | 354675.7 | 75.0000 | 4.00 | 5217.4 |
| 2002-04-01 | 356471.8 | 75.7333 | 4.00 | 5271.8 |
| 2002-07-01 | 359192.1 | 75.8000 | 4.00 | 4656.4 |
| 2002-10-01 | 362277.8 | 76.1667 | 4.00 | 3721.8 |
| 2003-01-01 | 364682.5 | 76.1333 | 3.83 | 3940.4 |
| 2003-04-01 | 368096.5 | 76.7667 | 3.75 | 3613.3 |
| 2003-07-01 | 371889.4 | 76.8000 | 3.50 | 4031.2 |
| 2003-10-01 | 375029.4 | 77.1667 | 3.67 | 4091.3 |
| 2004-01-01 | 377068.8 | 77.1667 | 3.92 | 4476.9 |
| 2004-04-01 | 378418.1 | 77.8000 | 4.25 | 4385.7 |
| 2004-07-01 | 379045.7 | 77.8333 | 4.67 | 4464.1 |
| 2004-10-01 | 380292.8 | 78.3333 | 4.75 | 4570.8 |
| 2005-01-01 | 383488.9 | 78.5333 | 4.75 | 4814.3 |
| 2005-04-01 | 388256.7 | 79.3000 | 4.75 | 4894.4 |
| 2005-07-01 | 392679.0 | 79.7000 | 4.58 | 5113.2 |
| 2005-10-01 | 398569.2 | 80.1000 | 4.50 | 5477.7 |
| 2006-01-01 | 400160.9 | 80.2000 | 4.50 | 5618.8 |
| 2006-04-01 | 401167.4 | 81.2333 | 4.50 | 5964.6 |
| 2006-07-01 | 401579.8 | 81.7333 | 4.67 | 5833.4 |
| 2006-10-01 | 403666.2 | 82.2667 | 4.92 | 5960.8 |
| 2007-01-01 | 407433.8 | 82.4333 | 5.25 | 6220.8 |
| 2007-04-01 | 409959.7 | 83.3000 | 5.42 | 6308 |
| 2007-07-01 | 413142.2 | 83.3333 | 5.75 | 6607.9 |
| 2007-10-01 | 415088.4 | 84.1333 | 5.67 | 6466.8 |
| 2008-01-01 | 417340.2 | 84.5333 | 5.33 | 6456.9 |
| 2008-04-01 | 415025.1 | 86.1000 | 5.00 | 5702.1 |
| 2008-07-01 | 408535.3 | 87.0667 | 5.00 | 5625.9 |
| 2008-10-01 | 400097.6 | 87.2333 | 3.17 | 4902.45 |
| 2009-01-01 | 393106.8 | 87.0333 | 1.00 | 4434.17 |
| 2009-04-01 | 392150.0 | 87.8333 | 0.50 | 3926.14 |

| 2009-07-01 | 392428.5 | 88.2000 | 0.50 | 4249.21 |
| 2009-10-01 | 393606.0 | 88.6333 | 0.50 | 5133.9 |
| 2010-01-01 | 396113.8 | 89.0667 | 0.50 | 5412.88 |
| 2010-04-01 | 400083.1 | 90.0667 | 0.50 | 5679.64 |
| 2010-07-01 | 402736.5 | 90.2667 | 0.50 | 4916.87 |
| 2010-10-01 | 402991.5 | 91.0667 | 0.50 | 5548.62 |
| 2011-01-01 | 405528.3 | 92.2333 | 0.50 | 5899.94 |
| 2011-04-01 | 405931.6 | 93.4333 | 0.50 | 5908.76 |
| 2011-07-01 | 407190.5 | 93.9667 | 0.50 | 5945.71 |
| 2011-10-01 | 407947.5 | 94.7333 | 0.50 | 5128.48 |
| 2012-01-01 | 410572.9 | 95.1000 | 0.50 | 5572.28 |
| 2012-04-01 | 410249.1 | 95.8000 | 0.50 | 5768.45 |
| 2012-07-01 | 415235.8 | 96.0667 | 0.50 | 5571.15 |
| 2012-10-01 | 414597.3 | 97.0333 | 0.50 | 5742.07 |
| 2013-01-01 | 417269.7 | 97.4333 | 0.50 | 5897.81 |
| 2013-04-01 | 419506.2 | 98.0667 | 0.50 | 6411.74 |
| 2013-07-01 | 423473.7 | 98.3667 | 0.50 | 6215.47 |
| 2013-10-01 | 425721.9 | 98.9333 | 0.50 | 6462.22 |
| 2014-01-01 | 428527.3 | 99.0333 | 0.50 | 6749.09 |
| 2014-04-01 | 431337.1 | 99.6667 | 0.50 | 6598.37 |
| 2014-07-01 | 433821.4 | 99.8333 | 0.50 | 6743.94 |
| 2014-10-01 | 436247.9 | 99.9667 | 0.50 | 6622.72 |
| 2015-01-01 | 438545.8 | 99.4333 | 0.50 | 6566.09 |
| 2015-04-01 | 441673.2 | 100.0333 | 0.50 | 6773.04 |
| 2015-07-01 | 443572.3 | 100.1667 | 0.50 | 6520.98 |
| 2015-10-01 | 446887.7 | 100.3333 | 0.50 | 6061.61 |
| 2016-01-01 | 447631.1 | 100.1333 | 0.50 | 6242.32 |
| 2016-04-01 | 450006.9 | 100.8000 | 0.50 | 6174.9 |
| 2016-07-01 | 452035.4 | 101.2000 | 0.33 | 6504.33 |
| 2016-10-01 | 454971.9 | 101.8667 | 0.25 | 6899.33 |
| 2017-01-01 | 457594.5 | 102.3000 | 0.25 | 7142.83 |
| 2017-04-01 | 458744.9 | 103.4000 | 0.25 | 7322.92 |
| 2017-07-01 | 460306.7 | 103.9333 | 0.25 | 7312.72 |
| 2017-10-01 | 462144.4 | 104.7000 | 0.5 | 7372.76 |
| 2018-01-01 | 462419.3 | 104.8333 | 0.5 | 7687.77 |
| 2018-04-01 | 464854.8 | 105.7667 | 0.5 | 7056.61 |
| 2018-07-01 | 467584.2 | 106.3333 | 0.5 | 7636.93 |
| 2018-10-01 | 468585.3 | 106.9000 | 0.75 | 7510.2 |
| 2019-01-01 | 471727.1 | 106.7333 | 0.75 | 6728.13 |
| 2019-04-01 | 470975.6 | 107.8000 | 0.75 | 7279.19 |
| 2019-07-01 | 473448.1 | 108.2333 | 0.75 | 7425.63 |
| 2019-10-01 | 473542.2 | 108.4333 | 0.75 | 7408.21 |
| 2020-01-01 | 464187.4 | 108.5000 | 0.25 | 7542.44 |

**Reference:** Federal Reserve Bank of St. Louis, Link: https://fred.stlouisfed.org

The original E-views software tables are placed below:

Table 1:

```
Equation: UNTITLED    Workfile: US DATA::Untitled\

Dependent Variable: IT
Method: Least Squares
Date: 06/27/20   Time: 23:54
Sample (adjusted): 1991Q1 2020Q1
Included observations: 117 after adjustments

Variable          Coefficient   Std. Error   t-Statistic   Prob.

INFLATION_GAP     0.906309      0.167241     5.419167      0.0000
OUTPUT_GAP        0.454512      0.179912     2.526311      0.0129
C                 2.451161      0.179028     13.69148      0.0000

R-squared              0.309850   Mean dependent var    2.712802
Adjusted R-squared     0.297742   S.D. dependent var    2.195087
S.E. of regression     1.839501   Akaike info criterion 4.082172
Sum squared resid      385.7490   Schwarz criterion     4.152997
Log likelihood        -235.8071   Hannan-Quinn criter.  4.110926
F-statistic            25.59074   Durbin-Watson stat    0.147860
Prob(F-statistic)      0.000000
```

Table 2:

Wald Test:
Equation: Untitled

| Test Statistic | Value | df | Probability |
|---|---|---|---|
| F-statistic | 3.142641 | (2, 114) | 0.0469 |
| Chi-square | 6.285282 | 2 | 0.0432 |

Null Hypothesis: C(1)=0.5, C(2)=0.5
Null Hypothesis Summary:

| Normalized Restriction (= 0) | Value | Std. Err. |
|---|---|---|
| -0.5 + C(1) | 0.406309 | 0.167241 |
| -0.5 + C(2) | -0.045488 | 0.179912 |

Restrictions are linear in coefficients.

Table 3:

```
Chow Breakpoint Test: 2003Q1
Null Hypothesis: No breaks at specified breakpoints
Varying regressors: All equation variables
Equation Sample: 1991Q1 2020Q1
```

| | | | |
|---|---|---|---|
| F-statistic | 56.80770 | Prob. F(3,111) | 0.0000 |
| Log likelihood ratio | 108.8485 | Prob. Chi-Square(3) | 0.0000 |
| Wald Statistic | 170.4231 | Prob. Chi-Square(3) | 0.0000 |

Table 4:

```
Chow Breakpoint Test: 2006Q1
Null Hypothesis: No breaks at specified breakpoints
Varying regressors: All equation variables
Equation Sample: 1991Q1 2020Q1
```

| | | | |
|---|---|---|---|
| F-statistic | 30.36553 | Prob. F(3,111) | 0.0000 |
| Log likelihood ratio | 70.10822 | Prob. Chi-Square(3) | 0.0000 |
| Wald Statistic | 91.09659 | Prob. Chi-Square(3) | 0.0000 |

Table 5:

```
Dependent Variable: IT
Method: Least Squares
Date: 06/27/20   Time: 23:59
Sample (adjusted): 1991Q2 2020Q1
Included observations: 116 after adjustments
```

| Variable | Coefficient | Std. Error | t-Statistic | Prob. |
|---|---|---|---|---|
| C | 2.367577 | 0.199163 | 11.88765 | 0.0000 |
| INFLATION_GAP | 0.842709 | 0.175769 | 4.794399 | 0.0000 |
| OUTPUT_GAP | 0.405135 | 0.202327 | 2.002372 | 0.0477 |
| S(-1) | 0.011777 | 0.012095 | 0.973657 | 0.3323 |

| | | | |
|---|---|---|---|
| R-squared | 0.303549 | Mean dependent var | 2.680753 |
| Adjusted R-squared | 0.284894 | S.D. dependent var | 2.176944 |
| S.E. of regression | 1.840910 | Akaike info criterion | 4.092271 |
| Sum squared resid | 379.5623 | Schwarz criterion | 4.187223 |
| Log likelihood | -233.3517 | Hannan-Quinn criter. | 4.130816 |
| F-statistic | 16.27178 | Durbin-Watson stat | 0.141748 |
| Prob(F-statistic) | 0.000000 | | |

Table 6:

```
Heteroskedasticity Test: White
Null hypothesis: Homoskedasticity

F-statistic              4.178675    Prob. F(9,107)         0.0001
Obs*R-squared           30.42807    Prob. Chi-Square(9)    0.0004
Scaled explained SS     12.92883    Prob. Chi-Square(9)    0.1659

Test Equation:
Dependent Variable: RESID^2
Method: Least Squares
Date: 06/28/20   Time: 00:01
Sample: 1991Q1 2020Q1
Included observations: 117

Variable                    Coefficient   Std. Error    t-Statistic    Prob.

C                            1.753718     0.468123      3.746273      0.0003
INFLATION_GAP^2              0.374279     0.154020      2.430060      0.0168
INFLATION_GAP*OUTPUT_G       0.424007     0.302290      1.402651      0.1636
INFLATION_GAP*S             -0.018188     0.020669     -0.880009      0.3808
INFLATION_GAP                0.421128     0.308246      1.366206      0.1747
OUTPUT_GAP^2                -0.487184     0.304197     -1.601540      0.1122
OUTPUT_GAP*S                -0.012456     0.026531     -0.469492      0.6397
OUTPUT_GAP                  -0.097270     0.496563     -0.195886      0.8451
S^2                          0.003006     0.001030      2.916840      0.0043
S                            0.061139     0.032539      1.878931      0.0630

R-squared              0.260069    Mean dependent var      3.263773
Adjusted R-squared     0.197832    S.D. dependent var      3.128591
S.E. of regression     2.802086    Akaike info criterion   4.980201
Sum squared resid    840.1305      Schwarz criterion       5.216284
Log likelihood      -281.3417      Hannan-Quinn criter.    5.076047
F-statistic            4.178675    Durbin-Watson stat      0.629901
Prob(F-statistic)      0.000121
```

Table 7:

```
Breusch-Godfrey Serial Correlation LM Test:
Null hypothesis: No serial correlation at up to 1 lag

F-statistic         650.9373    Prob. F(1,112)         0.0000
Obs*R-squared        99.82428   Prob. Chi-Square(1)    0.0000
```

Table 8:

```
Dependent Variable: IT
Method: Least Squares
Date: 06/28/20   Time: 00:02
Sample (adjusted): 1991Q1 2020Q1
Included observations: 117 after adjustments
HAC standard errors & covariance (Bartlett kernel, Newey-West fixed
    bandwidth = 5.0000)
```

| Variable | Coefficient | Std. Error | t-Statistic | Prob. |
|---|---|---|---|---|
| INFLATION_GAP | 0.891138 | 0.303483 | 2.936369 | 0.0040 |
| OUTPUT_GAP | 0.397858 | 0.310041 | 1.283241 | 0.2020 |
| S | 0.011997 | 0.020404 | 0.587946 | 0.5577 |
| C | 2.363806 | 0.307917 | 7.676756 | 0.0000 |

| | | | |
|---|---|---|---|
| R-squared | 0.316805 | Mean dependent var | 2.712802 |
| Adjusted R-squared | 0.298667 | S.D. dependent var | 2.195087 |
| S.E. of regression | 1.838289 | Akaike info criterion | 4.089137 |
| Sum squared resid | 381.8614 | Schwarz criterion | 4.183570 |
| Log likelihood | -235.2145 | Hannan-Quinn criter. | 4.127476 |
| F-statistic | 17.46647 | Durbin-Watson stat | 0.153153 |
| Prob(F-statistic) | 0.000000 | Wald F-statistic | 6.804897 |
| Prob(Wald F-statistic) | 0.000294 | | |

Table 9:

```
Dependent Variable: IT
Method: Generalized Method of Moments
Date: 06/28/20   Time: 00:04
Sample (adjusted): 1991Q3 2020Q1
Included observations: 115 after adjustments
Linear estimation with 1 weight update
Estimation weighting matrix: HAC (Bartlett kernel, Newey-West fixed
    bandwidth = 5.0000)
Standard errors & covariance computed using HAC weighting matrix
    (Bartlett kernel, Newey-West fixed bandwidth = 5.0000)
Instrument specification: INFLATION_GAP(-1) INFLATION_GAP(-2)
    OUTPUT_GAP(-1) OUTPUT_GAP(-2)
Constant added to instrument list
```

| Variable | Coefficient | Std. Error | t-Statistic | Prob. |
|---|---|---|---|---|
| C | 2.807578 | 0.448755 | 6.256373 | 0.0000 |
| INFLATION_GAP | 0.807066 | 0.424851 | 1.899645 | 0.0601 |
| OUTPUT_GAP | 0.931545 | 0.390469 | 2.385709 | 0.0187 |
| S | -0.052630 | 0.029490 | -1.784675 | 0.0770 |

| | | | |
|---|---|---|---|
| R-squared | 0.075589 | Mean dependent var | 2.653072 |
| Adjusted R-squared | 0.050605 | S.D. dependent var | 2.165870 |
| S.E. of regression | 2.110358 | Sum squared resid | 494.3506 |
| Durbin-Watson stat | 0.193156 | J-statistic | 3.683003 |
| Instrument rank | 5 | Prob(J-statistic) | 0.054970 |

Table 10:

```
Equation: UNTITLED   Workfile: UK DATA::Untitled\

Dependent Variable: IT
Method: Least Squares
Date: 06/27/20   Time: 22:57
Sample (adjusted): 1991Q1 2020Q1
Included observations: 117 after adjustments

Variable         Coefficient   Std. Error   t-Statistic   Prob.

C                3.458729      0.244900     14.12302      0.0000
INFLATION_GAP    1.256408      0.186942     6.720847      0.0000
OUTPUT_GAP       0.514578      0.226617     2.270698      0.0250

R-squared            0.292768    Mean dependent var    3.819715
Adjusted R-squared   0.280361    S.D. dependent var    3.033076
S.E. of regression   2.573006    Akaike info criterion 4.753333
Sum squared resid    754.7208    Schwarz criterion     4.824158
Log likelihood      -275.0700    Hannan-Quinn criter.  4.782087
F-statistic          23.59592    Durbin-Watson stat    0.048907
Prob(F-statistic)    0.000000
```

Table 11:

```
Equation: UNTITLED   Workfile: UK DATA::Untitled\

Wald Test:
Equation: Untitled

Test Statistic    Value       df         Probability

F-statistic       97.68412    (2, 114)   0.0000
Chi-square        195.3682    2          0.0000

Null Hypothesis: C(1)=0.5, C(2)=0.5
Null Hypothesis Summary:

Normalized Restriction (= 0)        Value       Std. Err.

-0.5 + C(1)                         2.958729    0.244900
-0.5 + C(2)                         0.756408    0.186942

Restrictions are linear in coefficients.
```

Table 12:

```
Equation: UNTITLED   Workfile: UK DATA::Untitled\

Chow Breakpoint Test: 2006Q1
Null Hypothesis: No breaks at specified breakpoints
Varying regressors: All equation variables
Equation Sample: 1991Q1 2020Q1

F-statistic            164.4359    Prob. F(3,111)         0.0000
Log likelihood ratio   198.2627    Prob. Chi-Square(3)    0.0000
Wald Statistic         493.3077    Prob. Chi-Square(3)    0.0000
```

Table 13:

```
Equation: UNTITLED   Workfile: UK DATA::Untitled\

Dependent Variable: IT
Method: Least Squares
Date: 06/27/20   Time: 23:11
Sample (adjusted): 1991Q2 2020Q1
Included observations: 116 after adjustments

Variable          Coefficient   Std. Error   t-Statistic    Prob.

C                  3.367140     0.252551     13.33250     0.0000
INFLATION_GAP      1.124711     0.209939      5.357315    0.0000
OUTPUT_GAP         0.388349     0.239484      1.621609    0.1077
S(-1)              0.025447     0.017705      1.437320    0.1534

R-squared          0.249156     Mean dependent var     3.740172
Adjusted R-squared 0.229044     S.D. dependent var     2.921103
S.E. of regression 2.564849     Akaike info criterion  4.755550
Sum squared resid  736.7863     Schwarz criterion      4.850502
Log likelihood    -271.8219     Hannan-Quinn criter.   4.794095
F-statistic        12.38848     Durbin-Watson stat     0.050940
Prob(F-statistic)  0.000000
```

Table 14:

| Heteroskedasticity Test: White | | | |
|---|---|---|---|
| Null hypothesis: Homoskedasticity | | | |
| F-statistic | 4.223505 | Prob. F(9,107) | 0.0001 |
| Obs*R-squared | 30.66894 | Prob. Chi-Square(9) | 0.0003 |
| Scaled explained SS | 12.41353 | Prob. Chi-Square(9) | 0.1910 |

Test Equation:
Dependent Variable: RESID^2
Method: Least Squares
Date: 06/19/20   Time: 16:14
Sample: 1991Q1 2020Q1
Included observations: 117

| Variable | Coefficient | Std. Error | t-Statistic | Prob. |
|---|---|---|---|---|
| C | 7.531529 | 0.734899 | 10.24839 | 0.0000 |
| INFLATION_GAP^2 | -0.632160 | 0.165757 | -3.813766 | 0.0002 |
| INFLATION_GAP*OUTPUT_G | -97.39981 | 58.53852 | -1.663858 | 0.0991 |
| INFLATION_GAP*S | -0.040076 | 0.032041 | -1.250785 | 0.2137 |
| INFLATION_GAP | 2.570281 | 0.701180 | 3.665652 | 0.0004 |
| OUTPUT_GAP^2 | -6222.799 | 2512.197 | -2.477034 | 0.0148 |
| OUTPUT_GAP*S | -2.273958 | 3.471984 | -0.654945 | 0.5139 |
| OUTPUT_GAP | -67.23506 | 56.31372 | -1.193937 | 0.2351 |
| S^2 | -0.001182 | 0.001990 | -0.594193 | 0.5536 |
| S | 0.013412 | 0.042030 | 0.319107 | 0.7503 |

| R-squared | 0.262128 | Mean dependent var | 6.373442 |
|---|---|---|---|
| Adjusted R-squared | 0.200064 | S.D. dependent var | 5.962916 |
| S.E. of regression | 5.333182 | Akaike info criterion | 6.267368 |
| Sum squared resid | 3043.382 | Schwarz criterion | 6.503451 |
| Log likelihood | -356.6410 | Hannan-Quinn criter. | 6.363215 |
| F-statistic | 4.223505 | Durbin-Watson stat | 0.464714 |
| Prob(F-statistic) | 0.000107 | | |

Table 15:

| Breusch-Godfrey Serial Correlation LM Test: | | | |
|---|---|---|---|
| Null hypothesis: No serial correlation at up to 1 lag | | | |
| F-statistic | 1702.731 | Prob. F(1,112) | 0.0000 |
| Obs*R-squared | 109.7791 | Prob. Chi-Square(1) | 0.0000 |

Table 16:

```
Dependent Variable: IT
Method: Least Squares
Date: 06/27/20   Time: 23:20
Sample (adjusted): 1991Q1 2020Q1
Included observations: 117 after adjustments
HAC standard errors & covariance (Bartlett kernel, Newey-West fixed
    bandwidth = 5.0000)
```

| Variable | Coefficient | Std. Error | t-Statistic | Prob. |
|---|---|---|---|---|
| INFLATION_GAP | 1.228155 | 0.313283 | 3.920280 | 0.0002 |
| OUTPUT_GAP | 0.464534 | 0.270615 | 1.716588 | 0.0888 |
| S | 0.019845 | 0.025339 | 0.783168 | 0.4352 |
| C | 3.383682 | 0.494236 | 6.846295 | 0.0000 |
| R-squared | 0.301228 | Mean dependent var | | 3.819715 |
| Adjusted R-squared | 0.282677 | S.D. dependent var | | 3.033076 |
| S.E. of regression | 2.568862 | Akaike info criterion | | 4.758393 |
| Sum squared resid | 745.6927 | Schwarz criterion | | 4.852826 |
| Log likelihood | -274.3660 | Hannan-Quinn criter. | | 4.796732 |
| F-statistic | 16.23744 | Durbin-Watson stat | | 0.057072 |
| Prob(F-statistic) | 0.000000 | Wald F-statistic | | 10.28791 |
| Prob(Wald F-statistic) | 0.000005 | | | |

Table 17:

```
Dependent Variable: IT
Method: Generalized Method of Moments
Date: 06/27/20   Time: 23:24
Sample (adjusted): 1991Q3 2020Q1
Included observations: 115 after adjustments
Linear estimation with 1 weight update
Estimation weighting matrix: HAC (Bartlett kernel, Newey-West fixed
    bandwidth = 5.0000)
Standard errors & covariance computed using HAC weighting matrix
    (Bartlett kernel, Newey-West fixed bandwidth = 5.0000)
Instrument specification: INFLATION_GAP(-1) INFLATION_GAP(-2)
    OUTPUT_GAP(-1) OUTPUT_GAP(-2)
Constant added to instrument list
```

| Variable | Coefficient | Std. Error | t-Statistic | Prob. |
|---|---|---|---|---|
| C | 3.627182 | 0.548552 | 6.612279 | 0.0000 |
| INFLATION_GAP | 1.132567 | 0.469443 | 2.412578 | 0.0175 |
| OUTPUT_GAP | 0.579766 | 0.375568 | 1.543707 | 0.1255 |
| S | -0.016528 | 0.057025 | -0.289837 | 0.7725 |
| R-squared | 0.160891 | Mean dependent var | | 3.672290 |
| Adjusted R-squared | 0.138213 | S.D. dependent var | | 2.840506 |
| S.E. of regression | 2.636913 | Sum squared resid | | 771.8172 |
| Durbin-Watson stat | 0.046020 | J-statistic | | 2.397154 |
| Instrument rank | 5 | Prob(J-statistic) | | 0.121556 |